\documentclass[prb,preprint]{revtex4-1} 

\usepackage{amsmath,amssymb,subfig,graphicx,hyperref}
\hypersetup{
  colorlinks   = true, 
  urlcolor     = blue, 
  linkcolor    = blue, 
  citecolor    = red   
}
\usepackage{color}

\newcommand{\x}{{\mathbf{x}}}
\renewcommand{\v}{{\mathbf{v}}}
\newcommand{\nhat}{{\hat{\mathbf{n}}}}
\newcommand{\vhat}{{\hat{\mathbf{v}}}}
\newcommand{\IR}{\ensuremath{\mathbb{R}}}

\begin{document}

\title{Is $n\sin\theta$ conserved along light path?}
\author{Mahdiyar Noorbala}
\affiliation{Department of Physics, University of Tehran, Tehran, Iran.  P.O.~Box 14395-547}
\author{Reza Sepehrinia}
\affiliation{Department of Physics, University of Tehran, Tehran, Iran.  P.O.~Box 14395-547}

\begin{abstract}
Snell's law states that the quantity $n\sin\theta$ is unchanged in refraction of light passing from one medium to another.  We inquire whether this is true in the general case where the speed of light varies continuously within a medium.  It turns out to be an instructive exercise in application of Snell's law and Fermat's principle.  It also provides good pedagogical problems in calculus of variations to deal with the subtleties of a variable domain of integration and inclusion of constraints.  The final result of these exercises is that, contrary to an initial expectation, the answer to the question in the title is negative.
\end{abstract}

\maketitle

\section{Introduction}\label{sec:intro}

\begin{figure}[b]
\centering
\subfloat[]{\includegraphics[scale=.4]{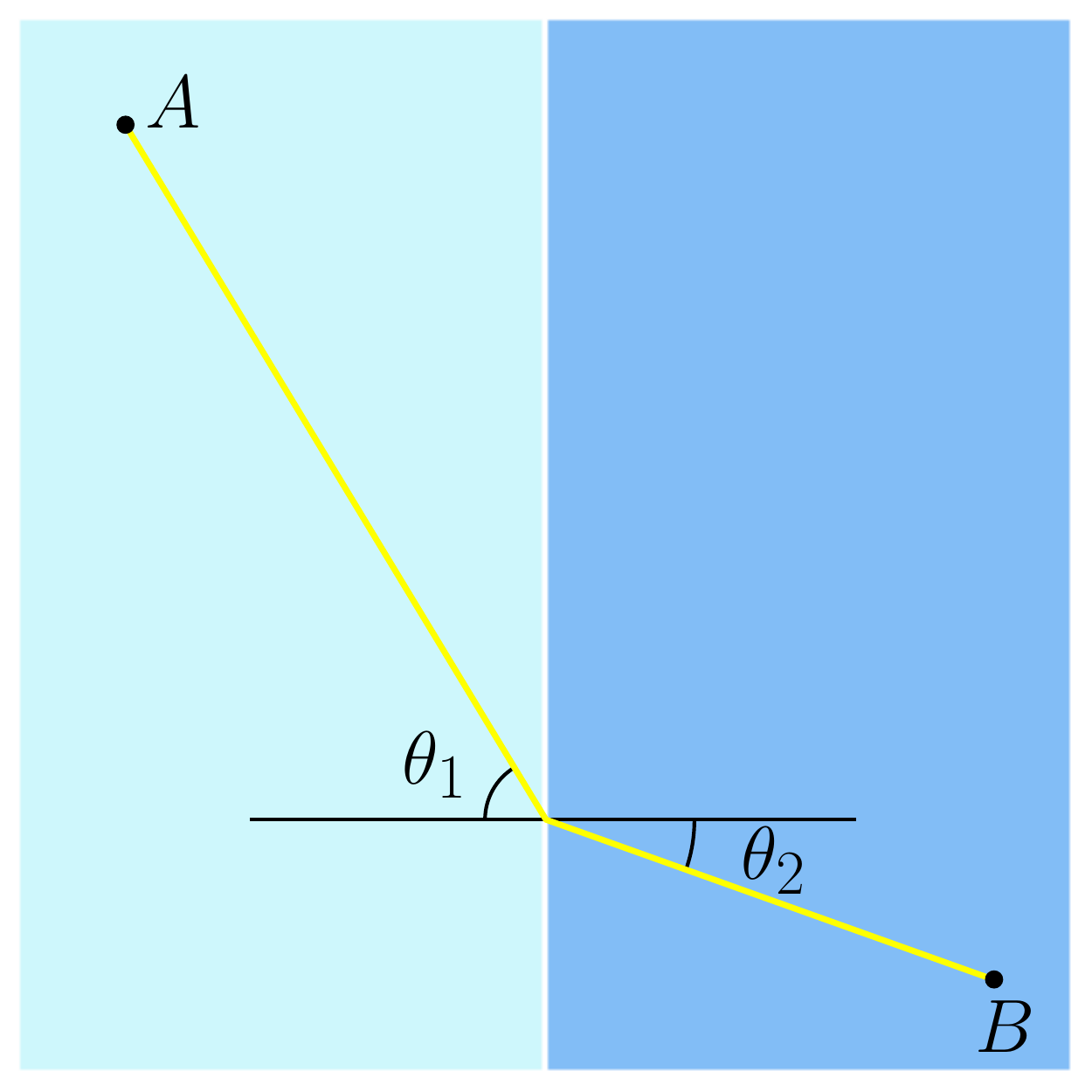}} \hspace{1cm}
\subfloat[]{\includegraphics[scale=.4]{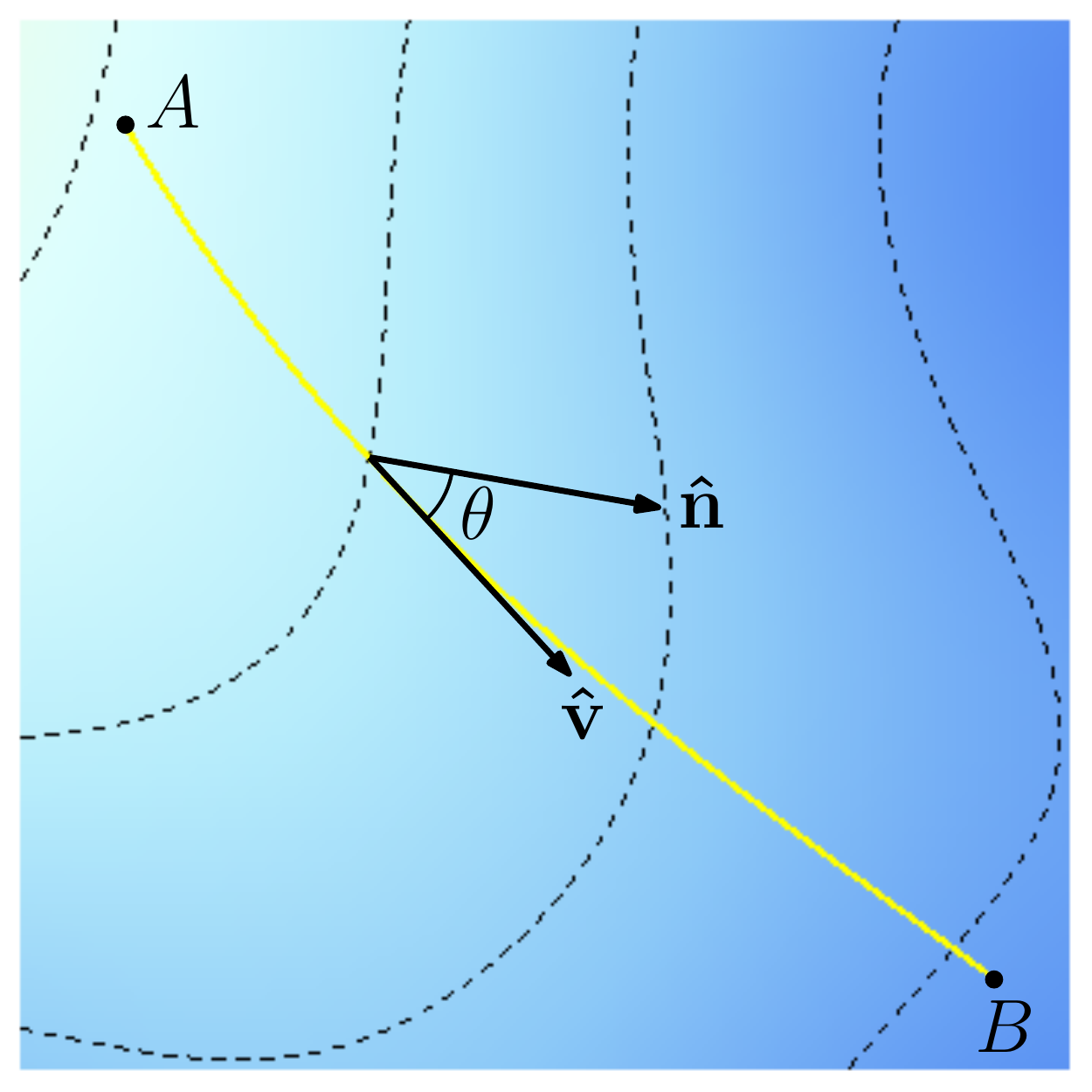}}
\caption{Trajectory of light from point $A$ to $B$.  The index of refraction $n$ is constant in each of the two media in (a), but varies in (b).  Darker color represents higher $n$ and light path (computed by solving the equations numerically) is shown in yellow.  Dashed lines denote surfaces of constant $n$ and $\nhat$ is orthogonal to them, while $\v$ is tangent to the path.}
\label{fig:1}
\end{figure}

A standard phenomenon in geometric optics is refraction of light across the boundary of two media with indices of refraction $n_1$ and $n_2$ (see Fig.~\ref{fig:1}(a)).  The well-known Snell's law states that
\begin{equation}\label{Snell}
n_1 \sin \theta_1 = n_2 \sin \theta_2,
\end{equation}
where $\theta_1$ is the incidence angle (the angle between the incident light ray and the normal to the boundary) and $\theta_2$ is the refraction angle (the angle between the refracted light ray and the normal to the boundary).  In fact this is also true when the boundary is a curved surface rather than a plane.\cite{geom-optics-approx}

A more complicated situation is when the refraction index varies continuously---rather than sharply as above---across space, as depicted in Fig.~\ref{fig:1}(b).  The main difference is the existence of a continuous gradient of index of refraction within the medium.  A host of interesting phenomena take place under this general situation.  Mirage is perhaps the most commonly known such phenomenon, in which temperature gradients above the hot ground induce variations in the index of refraction of air.  This in turn leads to a curved path of light that gives the observer the impression of the existence of a second virtual image.\cite{FLS,G,LGT,TOS}  Besides mirage, a wide range of applications is termed under gradient-index optics, where as the name suggests, variations in the index of refraction within a material are used for certain applications.  For example, in the core of a graded-index fiber the index of refraction decreases radially away from the axis.  This prevents rays from diverging away from the axis and hence reduces dispersion.  Another example is graded-index lenses.  Unlike conventional spherical lenses, a graded-index lens has a flat surface, but features a variable index of refraction in such a way that the desired focusing of light is achieved.

Snell's law, Eq.~\eqref{Snell}, tells us that whenever light crosses the interface between two media with different indices of refraction, the quantity $n\sin\theta$ is unchanged before and after refraction.  This reminds us of a conservation law.  Of course, Snell's law states this only for an abrupt change of the index of refraction as in Fig.~\ref{fig:1}(a).  One is then naturally inclined to ask if the same or a similar conservation law holds in the general situation of variable index of refraction.  In other words, we would like to ask if there is any quantity that is conserved on the light path in Fig.~\ref{fig:1}(b).

To answer this question, we may use the intuition from the previous situation to guess a path for the trajectory of light in the general case.  We can look at any point $\x$ on the trajectory of light in Fig.~\ref{fig:1}(b) and apply Snell's law for the infinitesimal region surrounding it.  Now the gradient $\nabla n$ plays the role of the normal to the boundary separating the two regions with infinitesimally different values of $n$.  So the expectation will be that the quantity $n(\x)\sin\theta(\x)$ is conserved along the path if we interpret $n(\x)$ as the local refraction index at point $\x$, and $\theta(\x)$ as the angle between the path and $\nabla n(\x)$.

Note that we have not found an explicit equation for the path.  All we have done is to guess a candidate conserved quantity along the path, based on the simple form of Snell's law.  In the next section we find an explicit equation for the path and check whether $n(\x)\sin\theta(\x)$ is indeed conserved along the path.

\section{Light Path}\label{sec:light-path}

We use Fermat's principle to obtain the path of light in a medium with variable index of refraction $n(\x)$.  Fermat's principle states that the path of light between two given points is the one which minimizes the travel time.  If the speed of light is constant in the medium containing the two points, the obvious conclusion is that light will travel on the straight line that joins the two points.  A more complex situation is where the medium is comprised of two parts with constant but different speeds of light.  One can then show from Fermat's principle, using basic algebra, that the light path will be straight in each medium with a sharp edge at the boundary and the angle between the two paths obeys Snell's law \eqref{Snell}.  If, however, the properties of the medium are such that the speed of light varies from point to point, then a more elaborate variational calculation is required as follows.

Suppose $\gamma$ is a path that joins points $\x_1$ and $\x_2$.  We can parameterize $\gamma$, i.e., write
\begin{equation}
\gamma = \{ \x(s) | 0 \leq s \leq 1 \text{ and } \x(0)=\x_1, \x(1)=\x_2 \},
\end{equation}
so that the function $\x(s)$ spans $\gamma$ as the parameter $s$ varies between 0 and 1.  Our aim is to find an equation for $\x(s)$ by requiring that the total travel time of light be minimized.  It is worthwhile to note, ahead of any computation, that although we expect the light path $\gamma$ to be unique, the function $\x(s)$ is not unique.  As a simple example, both functions $\x(s)=s\v$ and $\x(s)=s^2\v$ represent a line parallel to the constant vector $\v$ and hence describe the same path $\gamma$.

For a linear isotropic medium the speed of light at a point is given by $c/n(\x)$, where $c$ is the speed of light in vacuum and $n(\x)$ is the refraction index at that point.  Noting that infinitesimal length element along $\gamma$ is just the magnitude of $d\x$, we can write the total travel time as
\begin{equation}\label{s01-param-action}
T = \frac1c \int_{0}^{1} n(\x(s)) \sqrt{\dot\x(s) \cdot \dot\x(s)} ds,
\end{equation}
where dot denotes $d/ds$.  Fermat's principle prescribes that the actual path $\gamma$ that is followed by light minimizes this integral.

Now this is a standard variational problem, so we can use the toolkit of calculus of variations to find the light path.\cite{G,LGT,LP,S,B}  We have chosen a simple parameterization of $\gamma$ to make it an easy variational problem.  But it is instructive to work with other parameterizations that require a more careful treatment.  The interested reader is encouraged to read Appendix \ref{app} for details, and Refs.~\onlinecite{LP,S} for a more advanced treatment.  For the present case, we note that the integral in Eq.~\eqref{s01-param-action} is a special case of Eq.~\eqref{action} of Appendix \ref{app}, and thus according to the rules of calculus of variations, $\x(s)$ must satisfy the Euler-Lagrange equation \eqref{EL}, which now reads
\begin{equation}
\frac{d}{ds} \frac{\partial}{\partial \dot x_i}  \left[ n(\x(s)) \sqrt{\dot\x(s) \cdot \dot\x(s)} \right] = \frac{\partial}{\partial x_i} \left[ n(\x(s)) \sqrt{\dot\x(s) \cdot \dot\x(s)} \right], \qquad i=1,2,3.
\end{equation}
Performing $\partial/\partial \dot x_i$ and $\partial/\partial x_i$ derivatives, we obtain the result
\begin{equation}\label{eq1-for-v}
\frac{d}{ds} \left( n \vhat \right) = v \nabla n,
\end{equation}
where we have defined $\v(s) := \dot\x(s)$ (as usual, $v$ and $\vhat$ are the magnitude and direction of $\v$) and all function arguments are suppressed.  The gradient on $n$ appears because of differentiation with respect to $x_i$.  If we also carry out $d/ds$ we need to take into account the $s$-dependence of $n$ through its $\x$-dependence: $\frac{dn}{ds} = \frac{\partial n}{\partial x_i} \frac{dx_i}{ds} = \v \cdot \nabla n$.  So we may write Eq.~\eqref{eq1-for-v} equivalently as
\begin{equation}\label{eq2-for-v}
n\frac{d\vhat}{ds} = v \nabla n - \v \left( \vhat\cdot\nabla n \right).
\end{equation}

Equations~\eqref{eq1-for-v} and \eqref{eq2-for-v} are our equations for the path of light $\x(s)$.  It is indeed hard to solve any of these equations, since for generic $n(\x)$ they are non-linear (both in $\x$ and $\dot\x$) second order differential equations.  Even the planar trajectories when the incident ray is initially tangent or normal to the surfaces of constant $n$ are not trivial. 
Also note that if $\x(s)$ is any solution to \eqref{eq1-for-v} then $\x(\bar s(s))$ is also a solution, for any monotonic function $\bar s(s)$ satisfying $\bar s(0)=0$ and $\bar s(1)=1$.  This reflects the non-uniqueness we mentioned above.

Now that we have the equation for the path at our disposal, we turn to the question of conservation of $n\sin\theta$.  We first note that $\theta$ is the angle between the path and $\nabla n$.  Thus defining the unit vector $\nhat:=\nabla n / |\nabla n|$, we have
\begin{equation}
|\sin\theta| = |\nhat \times \vhat|.
\end{equation}
We can now compute
\begin{equation}
\begin{aligned}
\frac{d}{ds} (n\sin\theta)^2 &= \frac{d}{ds} [(\nhat \times n\vhat) \cdot (\nhat \times n\vhat)] \\
&= 2 (\nhat \times n\vhat) \cdot \left(\frac{d\nhat}{ds} \times n\vhat + \nhat \times \frac{d}{ds} (n\vhat) \right).
\end{aligned}
\end{equation}
Now $\nhat \times d(n\vhat)/ds$ vanishes by means of Eq.~\eqref{eq1-for-v}.  Then we can use $(A\times B)\cdot C=(C\times A)\cdot B$ (with $A=\nhat$, $B=n\vhat$, and $C=d\nhat/ds\times n\vhat$) to obtain
\begin{equation}
\frac{d}{ds} (n\sin\theta)^2 = 2 \left[ \left(\frac{d\nhat}{ds} \times n\vhat \right) \times \nhat \right] \cdot n\vhat.
\end{equation}
If we apply $(A\times B)\times C=(A\cdot C)B-(B\cdot C)A$ (with $A=d\nhat/ds$, $B=n\vhat$, and $C=\nhat$) to the quantity inside the brackets, we get
\begin{equation}
\begin{aligned}\label{noncons}
\frac{d}{ds} (n\sin\theta)^2 &= 2 \left[ \left( \frac{d\nhat}{ds} \cdot \nhat \right) n\vhat - \left( n\vhat \cdot \nhat \right) \frac{d\nhat}{ds} \right] \cdot n\vhat \\
&= -2n^2 (\nhat \cdot \vhat) \left(\frac{d\nhat}{ds} \cdot \vhat \right),
\end{aligned}
\end{equation}
where we have used $\nhat \cdot d\nhat=0$, since $\nhat$ is a unit vector and $d(\nhat\cdot\nhat)=0$.  This is in marked contrast with our expectation that $n\sin\theta$ should be conserved.  To better appreciate the difference, let us simplify this result when everything happens in a plane, i.e., $\nhat$, $\vhat$ and $d\nhat/ds$ are coplanar.  Let $\theta$ measure the angle from $\nhat$ to $\vhat$ in the counterclockwise direction (so for example, $\theta<0$ in Fig.~\ref{fig:1}(b)).  Then, since $\nhat\perp d\nhat/ds$, the angle between $d\nhat/ds$ and $\vhat$ is $\pi/2\pm\theta$, where the $\pm$ sign applies when $d\nhat/ds$ is obtained by a $\pm90$-degree clockwise rotation of $\nhat$.  Then Eq.~\eqref{noncons} becomes
\begin{equation}
2 (n\sin\theta) \frac{d}{ds} (n\sin\theta) = -2n^2 \cos\theta \cos(\frac{\pi}{2}\pm\theta) \left| \frac{d\nhat}{ds} \right|,
\end{equation}
or
\begin{equation}\label{noncons-planar}
\frac{d}{ds} (n\sin\theta) = \pm n \cos\theta \left| \frac{d\nhat}{ds} \right|.
\end{equation}
We see that $n\sin\theta$ is not conserved, unless $\nabla n$ is a unidirectional vector field or the incident light is tangent to the surface of constant $n$.

We need to understand what was wrong in our original expectation.

\section{Comparison with Snell's Law}

We now try to revisit the na\"ive argument that led to the conservation of $n\sin\theta$ and justify its nonconservations using simple arguments based on Snell's law.

\begin{figure}
\centering
\includegraphics[scale=1]{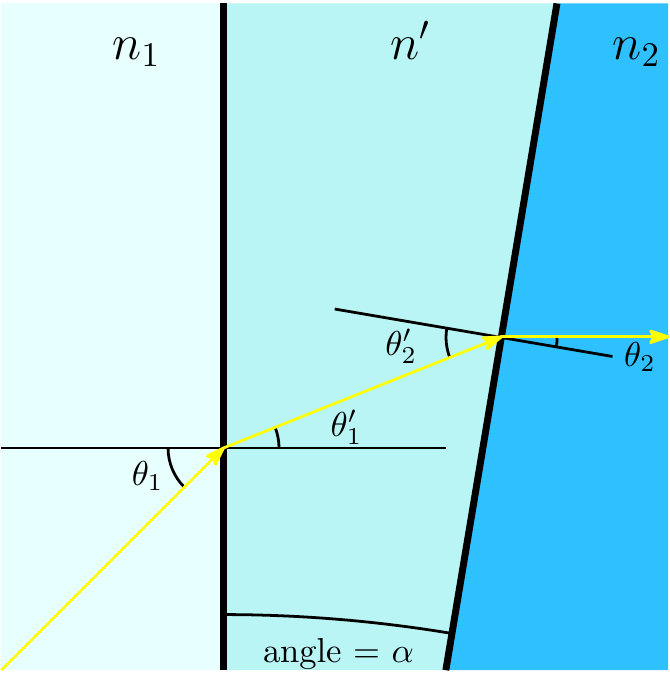}
\caption{Light path in a medium with variable index of refraction and non-parallel surfaces of constant $n$.  The infinitesimal intermediate region with index $n'$ is magnified.}
\label{fig:2}
\end{figure}

The key point in resolving the problem is to accurately depict the regions of constant $n$ in the neighborhood of the point of interest.  This is done in Fig.~\ref{fig:2}, where a planar trajectory of light is shown passing from a region with index of refraction $n_1$ to an infinitesimally nearby one with $n_2$.  The surfaces of constant $n$, and hence the gradients $\nabla n$, are not parallel.  This means that the intermediate region with index of refraction $n_1<n'<n_2$ has non-parallel boundaries which make an infinitesimal angle $\alpha$.  The incident (refraction) angle is $\theta_1$ ($\theta_2$) as measured from the normal to the left (right) boundary of the $n'$ region.  We want to compare $n_1\sin\theta_1$ with $n_2\sin\theta_2$.

We need to related $\theta_2$ to $\theta_1$ through the intermediate angles $\theta'_1$ and $\theta'_2$.  We can write Snell's law for the left and right boundary, respectively:
\begin{equation}
n_1\sin\theta_1 = n'\sin\theta'_1, \qquad n_2\sin\theta_2 = n'\sin\theta'_2.
\end{equation}
The main point is that $\theta'_1\neq\theta'_2$; if they were equal, we would readily obtain $n_1\sin\theta_1 = n_2\sin\theta_2$ from the above equation.  Instead, from the geometry of Fig.~\ref{fig:2} we have $\theta'_2 = \theta'_1 + \alpha$.  We can now Taylor expand $\sin(\theta'_1+\alpha)$ to the leading order in $\alpha$ to obtain
\begin{equation}
n_2\sin\theta_2 - n_1\sin\theta_1 = \alpha n' \cos \theta'_1 + {\cal O}(\alpha^2).
\end{equation}
This is precisely the same as the analytic result \eqref{noncons-planar}, once we note that $\alpha = |d\nhat/ds|$ and the situation in Fig.~\ref{fig:2} corresponds to the $+$ sign of Eq.~\eqref{noncons-planar}.

We observe that the falsehood of our original expectation about the conservation of $n\sin\theta$ is related to ignoring the tilt of the gradient vector $\nabla n$.  Once this is carefully taken into account, there is no discrepancy.

\section{What Is Conserved?}

We showed that our na\"ive guess about the conservation of $n\sin\theta$ is incorrect.  The next natural question to ask is ``what is conserved then?''  A similar question can be asked in classical mechanics of particles.  In that case, the conservation laws are known to be consequences of symmetries.  So it pays if we view our problem as the motion of a particle in a potential.  To this end, it is easiest to work with a length parameterization where $s=\ell$ is the length of the curve and $v=|\dot\x|=1$.  It then follows that Eq.~\eqref{eq1-for-v} becomes
\begin{equation}
\frac{d}{d\ell}(n\vhat) = \nabla n
\end{equation}
(compare with Eq.~\eqref{eq1-for-v-ell} of Appendix~\ref{app:length}).  This can now be compared with the equation
\begin{equation}
\frac{d}{dt}{\bf p} = -\nabla V
\end{equation}
for the trajectory of a particle under the influence of the potential $V(\x)$.  

We know from classical mechanics that if $V(\x)$ has a symmetry, then there exists a quantity that is conserved in time $t$.  So we conclude in analogy that if $n(\x)$ has a symmetry, then there exists a quantity that is conserved in $\ell$, that is, along the path.  As an example, if $V(\x)$ has translational symmetry along some given direction (e.g., is independent of the horizontal distance $x$), we know that the conserved quantity is the component of momentum $\bf p$ along that direction (e.g., $p_x$).  We can thus draw the conclusion that when $n(\x)$ has translational symmetry in the direction perpendicular to $\nabla n$ (i.e., the surfaces of constant $n$ are parallel flat planes), then the conserved quantity is the component of $n\vhat$ along $\nabla n$.  This is precisely the statement of conservation of $n\sin\theta$.

This analogy tells us that the conservation of $n\sin\theta$ is related to the symmetry of $n(\x)$.  But it also tells us that when $n(\x)$ has no symmetry, we should not expect anything to be conserved.

At this point it is worthwhile to mention a caveat in our analysis when it comes to real world applications.  As indicated in footnote~\onlinecite{geom-optics-approx}, in this paper we are solely concerned with geometrical optics, where the wavelengths of rays are much smaller than the scales of the problem.  So we have ignored the phase of the wave altogether.  However, there are situations in which an abrupt phase change takes place somewhere along the path that can dramatically affect the analysis, even though the wavelengths are still negligible.  This is because the actual form of Fermat's law (also called generalized Fermat's law) is about extremizng the \textit{phase change} of light, rather than its \textit{travel time}.  When there is no change of phase due to external agents, the only way the phase $\Phi$ changes between two points is through $\delta\Phi={\bf k}\cdot\delta\x\propto\delta t$; so extremizing the phase amounts to extremizing the time.  But in general we have to take into account the effect of other forms of phase change if they are present.  We are not going to discuss the generalized Fermat's law any further here, except to quote the generalized form of the refraction law for the case where the abrupt phase change is imparted to the wave in passing through a flat interface like Fig.~\ref{fig:1}(a)
\begin{equation}
n_1\sin\theta_1 - n_2\sin\theta_2 = \frac{\lambda_0}{2\pi} \frac{d\Phi}{dy},
\end{equation}
where $\lambda_0$ is the wavelength of the light in vacuum, and $y$ is the vertical direction along the interface.  We see that the effect of a gradient of $\Phi$ is to violate the conservation of $n\sin\theta$ even in the flat interface case.  For detailed derivation and application, see Refs.~\onlinecite{YGKATCG,WSHL}.

\section{Summary}

We began with a heuristic but na\"ive argument that led to the conclusion that $n\sin\theta$ must be conserved along light path in a medium with variable speed of light.  We then analyzed the problem using Fermat's principle and calculus of variations.  (Several other tricky versions of the variational problem were discussed in Appendix~\ref{app}.  They deal with the subtleties of a variable domain of integration and inclusion of constraints.)  Comparison with the analytic result revealed that our original expectation was wrong.  We finally returned to our heuristic argument and identified the problem.  We calculated the amount of nonconservation of $n\sin\theta$ and found exactly the same answer as the one we obtained from the variational approach.

\appendix

\section{Variation with Other Parameterizations}\label{app}

In this appendix we find the equation for the path of light using different parameterizations that make the variational problem harder.  It is only of pedagogical value, since we already know that Eqs.~\eqref{eq1-for-v} and \eqref{eq2-for-v} are the answer.  Indeed it turns to be a valuable exercise in variational techniques and so we present the analysis in this appendix.

Let us briefly review the basics results of the variational method.  The computational framework is easy to follow once a functional (function of a function)
\begin{equation}\label{action}
I[x] = \int_{t_1}^{t_2} L(x(t), \dot x(t)) dt
\end{equation}
is identified, which is supposed to be extremized (stationary) under variations of the function $x$.  The main result of calculus of variations is that $I$ is stationary if the Euler-Lagrange equation,
\begin{equation}\label{EL}
\frac{\partial L}{\partial x} - \frac{d}{dt} \left( \frac{\partial L}{\partial \dot x} \right) = 0,
\end{equation}
is satisfied.  When there are several functions $x$, we have one such equation for each $x$.  We often refer to $I$ as ``functional'' or ``action'', and to $L$ as ``integrand'' or ``Lagrangian''.  The Euler-Lagrange equations are also called ``equations of motion''.

This much of the formalism is sufficient for most applications.  Indeed this was all we used in Section \ref{sec:light-path}.  But in what follows we will be dealing with situations with constraints.  In such problems $I$ is supposed to be extremized subject to a constraint $J[x]=0$, which is itself a functional of $x$.  Then one forms the new functional $K=I+\mu J$, where $\mu$ is a constant known as the Lagrange multiplier.  The Euler-Lagrange equations suitable for the constrained problem are then exactly as Eq.~\eqref{EL}, except that $I$ is changed to $K$.  Finally, when the constraint is of algebraic form---like $j(x(t), \dot x(t)) = 0$ for all $t$---then $\mu(t)$ is an infinite set of Lagrange multipliers---one for each $t$---and the new functional $K=I+\int_{t_1}^{t_2} \mu jdt$ is the object to satisfy Eq.\eqref{EL}.\cite{nonholonomic}  All of these straightforwardly generalize to the case of several functions and several constraints.  In all cases, the $\mu$s are to be regarded as independent functions just like the $x$s.

There are two parameterizations that we are going to discuss in this appendix: parameterization of the path by arc length and by time.  In the first case, the integral to minimize is
\begin{equation}
T = \frac1c \int n(\x(\ell)) d\ell,
\end{equation}
while in the second case
\begin{equation}
T = \int dt.
\end{equation}
As we will see, there are constraints to be appended to these integrals.  But before doing that, there is another subtlety that we should discuss first.  It is related to the fact that in these integrals the domain of integration is not a priori known and fixed.  (For example, if we knew the domain of integration in parameterization with time, then we already knew $T$.)  This is very important, since the standard results we quoted above are usually derived under the assumption that there is a fixed domain of integration and the variations vanish at its boundary.\cite{domain-vs-endpoint}  So in the rest of this appendix we first address the issue of variable domain in Subsection \ref{app:var-domain}, and then turn to the length and time parameterizations in Subsections \ref{app:length} and \ref{app:time}, respectively.

\subsection{Variation with Variable Domain}\label{app:var-domain}

In Section \ref{sec:light-path} our parametererization was over a fixed domain, namely: $0\leq s\leq 1$.  This meant fixed lower and upper limits ($0$ and $1$) in the integral \eqref{s01-param-action} and hence we were eligible to use the Euler-Lagrange equations.  However, we may equally well describe the path connecting $\x_1$ and $\x_2$ by a function $\x(s)$ on an arbitrary domain $\sigma_1\leq s\leq\sigma_2$.  The travel time is still given by the same integral except for the lower and upper limits of integration:
\begin{equation}\label{s-param-action}
T = \frac1c \int_{\sigma_1}^{\sigma_2} n(\x(s)) \sqrt{\dot\x(s) \cdot \dot\x(s)} ds.
\end{equation}
We wish to find such general parameterizations of the path that minimize $T$.  This appears to be a harder problem in calculus of variations, since: (i) we have to search among a larger class of functions; and (ii) the domain of integration is not fixed as before, but variable.

Our search will be among functions
\begin{equation}
\x:[\sigma_1,\sigma_2]\to\IR^3, \qquad \text{such that $\x(\sigma_1)=\x_1$ and $\x(\sigma_2)=\x_2$.}
\end{equation}
This is to be contrasted with our previous search scope, $\x:[0,1]\to\IR^3$.  Under a variation $\delta\x$ in the function and $\delta\sigma_{1,2}$ in the domain, the boundary conditions imply
\begin{equation}
(\x+\delta\x)\big|_{\sigma_1+\delta\sigma_1} = \x_1, \qquad (\x+\delta\x)\big|_{\sigma_2+\delta\sigma_2} = \x_2,
\end{equation}
which means
\begin{equation}
\delta\x(\sigma_1) = -\dot\x(\sigma_1) \delta\sigma_1, \qquad \delta\x(\sigma_2) = -\dot\x(\sigma_2) \delta\sigma_2.
\end{equation}
Now the variation of $T=\int_{\sigma_1}^{\sigma_2} Lds$ becomes
\begin{equation}\label{dT-variable-domain}
\delta T = \int_{\sigma_1}^{\sigma_2} \left[ \frac{\partial L}{\partial x_i} - \frac{d}{ds} \frac{\partial L}{\partial \dot x_i} \right] \delta x_i ds + \left[ L - \dot x_i \frac{\partial L}{\partial \dot x_i} \right]_{\sigma_2} \delta\sigma_2 - \left[ L - \dot x_i \frac{\partial L}{\partial \dot x_i} \right]_{\sigma_1} \delta\sigma_1,
\end{equation}
where summation over $i$ is implied.  By inspecting \eqref{s-param-action} it is clear that $L$ is a homogeneous function of $\dot\x$ and hence the last two terms in $\delta T$ vanish.  Thus we are again left with the standard Euler-Lagrange equation and the result Eq.~\eqref{eq1-for-v}.  Again we observe the non-uniqueness in the solutions:  If $\x(s):[\sigma_1,\sigma_2]\to\IR^3$ satisfies Eq.~\eqref{eq1-for-v}, then
\begin{equation}\label{reparam}
\x(\bar s(s)):[\bar\sigma_1,\bar\sigma_2]\to\IR^3
\end{equation}
is also a solution, for any monotonic function $\bar s(s)$ satisfying $\bar s(\sigma_1)=\bar\sigma_1$ and $\bar s(\sigma_2)=\bar\sigma_2$.

We could have guessed that the Euler-Lagrange equations remain untouched:  A function that minimizes the integral among all functions with variable domains, also does so among the smaller class of functions which share its own domain.  So in general, the Euler-Lagrange equations provide a necessary condition for variable-domain problems.  The boundary terms that appear in Eq.~\eqref{dT-variable-domain} may provide additional conditions, although in our case they don't.

\subsection{Parameterization with Length}\label{app:length}

We are now in a position to perform the variational analysis with length parameterization.  So we choose to set the arc length $\ell$ run from zero to $\lambda$, the total length of the path, and work with $\x(\ell)$.  This choice corresponds to the constraint $| d\x/d\ell | = 1$, to handle which we need to employ a Lagrange multiplier function $\mu(\ell)$.  Therefore, the quantity to minimize is
\begin{equation}\label{l-param-action}
\frac1c \int_{0}^{\lambda} n(\x(\ell)) d\ell + \int_{0}^{\lambda} \mu(\ell) \left[ \dot\x(\ell) \cdot \dot\x(\ell) - 1 \right] d\ell,
\end{equation}
where dot now means $d/d\ell$.  Again we have a variable domain of integration, since we don't know the length $\lambda$ of the desired trajectory before solving for it.  Requiring $\x(\lambda)=\x_2$ implies
\begin{equation}
\delta\x(\lambda) = -\dot\x(\lambda) \delta\lambda,
\end{equation}
while variation of Eq.~\eqref{l-param-action} yields
\begin{multline}\label{dT-l-param}
\int_{0}^{\lambda} \left[ \frac1c \frac{\partial n}{\partial x_i} - \frac{d}{d\ell} \left( 2\mu\dot x_i \right) \right] \delta x_i d\ell +
\int_{0}^{\lambda} \left[ |\dot\x|^2 - 1 \right] \delta \mu d\ell \\
+ \left[ \frac1c n + \mu \left( |\dot\x|^2 - 1 \right) - 2\mu |\dot\x|^2 \right]_{\lambda} \delta\lambda.
\end{multline}

We can now treat the variations $\delta x_i$, $\delta \mu$, and $\delta \lambda$ independently and set their coefficients to zero.  The coefficient of $\delta \mu$ is nothing but the constraint $|\dot\x|=1$ we imposed.  The coefficient of $\delta x_i$ gives the usual Euler-Lagrange equation---as we saw earlier this is always a necessary condition in variable domain problems---which reads
\begin{equation}\label{EL-l-param}
\frac1c \frac{\partial n}{\partial x_i} = \frac{d}{d\ell} \left( 2\mu\dot x_i \right).
\end{equation}
Now this equation involves the unknown function $\mu$, so we need to use $|\dot\x|=1$ to get rid of it.  In order to do so, we multiply our equation through by $\dot x_i$ and sum over $i$.  We further note that $\dot{x}_i\ddot{x}_i=0$ and $\dot{n}=\dot{x}_i \partial n/\partial x_i$, to obtain
\begin{equation}
\frac{d}{d\ell} \left( \frac1c n - 2\mu \right) = 0.
\end{equation}
This equation almost fixes $\mu$, except for an unknown constant of integration.  To proceed, we recall that the coefficient of $\delta \lambda$ in the variation \eqref{dT-l-param} must also vanish, yielding
\begin{equation}
\frac1c n(\x(\lambda)) = 2 \mu(\lambda).
\end{equation}
We observe that, unlike the situation with Eq.~\eqref{dT-variable-domain}, the boundary terms do provide additional conditions, which in fact turn out to be quite useful and fix our unknown constant of integration.  Thus we obtain
\begin{equation}
\frac1c n(\x(\ell)) = 2 \mu(\ell), \qquad \forall \ell \in [0,\lambda].
\end{equation}
Substituting $\mu$ from this equation into Eq.~\eqref{EL-l-param}, we get
\begin{equation}\label{eq1-for-v-ell}
\frac{\partial n}{\partial x_i} = \frac{d}{d\ell} \left( n\dot x_i \right).
\end{equation}
It is easy to see that this is the same as Eq.~\eqref{eq1-for-v} that we found for the generic parameterization ($s=\ell$ and $v=1$ here).

\subsection{Parameterization with Time}\label{app:time}

As our second special case we now consider parameterization with time.  Just as we used $\ell$ in the previous subsection, we are now going to employ $t$ running from zero to $\tau$, the total time of travel, and work with $\x(t)$.  Our new constraint is $| d\x/dt | = c/n$ and again we need a Lagrange multiplier function $\mu(t)$.  We now want to minimize
\begin{equation}\label{t-param-action}
\int_{0}^{\tau} dt + \int_{0}^{\tau} \mu(t) \left[ \dot\x(t) \cdot \dot\x(t) - \left( \frac{c}{n(\x(t))} \right)^2 \right] dt,
\end{equation}
where dot now means $d/dt$.  Just like before, we have a variable domain of integration: we don't know the travel time $\tau$ of the desired trajectory before solving for it.  Requiring $\x(\tau)=\x_2$ implies
\begin{equation}
\delta\x(\tau) = -\dot\x(\tau) \delta\tau.
\end{equation}
If we now vary Eq.~\eqref{t-param-action}, we find
\begin{multline}\label{dT-t-param}
\int_{0}^{\tau} \left[ \frac{2c^2\mu}{n^3} \frac{\partial n}{\partial x_i} - \frac{d}{dt} \left( 2\mu\dot x_i \right) \right] \delta x_i dt +
\int_{0}^{\tau} \left[ |\dot\x|^2 - \left( \frac{c}{n} \right)^2 \right] \delta \mu dt \\
+\left[ 1 + \mu \left( |\dot\x|^2 - \left( \frac{c}{n} \right)^2 \right) - 2\mu |\dot\x|^2 \right]_{\tau} \delta\tau.
\end{multline}
The coefficient of $\delta \mu$ gives our constraint and that of $\delta x_i$ gives the standard Euler-Lagrange equation:
\begin{equation}\label{EL-t-param}
\frac{2c^2\mu}{n^3} \frac{\partial n}{\partial x_i} = \frac{d}{dt} \left( 2\mu\dot x_i \right).
\end{equation}
To eliminate $\mu$ we invoke the constraint, and in steps parallel to what we did before, multiply the above equation by $\dot x_i$ and sum over $i$.  Again we use $\dot{n}=\dot{x}_i \partial n/\partial x_i$, but this time we note that $\dot{x}_i\ddot{x}_i = \frac12 d(c/n)^2/dt$.  We obtain
\begin{equation}
\frac{d}{dt} \left( \frac{2c^2\mu}{n^2} \right) = 0.
\end{equation}
As before, this equation almost fixes $\mu$, except for an unknown constant of integration.  On the other hand, the coefficient of $\delta \tau$ in the variation \eqref{dT-t-param} must also vanish, yielding
\begin{equation}
2\mu(\tau) = \frac{1}{|\dot\x(\tau)|^2} = \frac{n^2(\x(\tau))}{c^2}.
\end{equation}
This fixes the unknown constant and we find
\begin{equation}
2\mu(t) = \frac{1}{|\dot\x(t)|^2} = \frac{n^2(\x(t))}{c^2}, \qquad \forall t \in [0,\tau].
\end{equation}
Substituting $\mu$ from this equation into Eq.~\eqref{EL-t-param}, we get
\begin{equation}
\frac{1}{n} \frac{\partial n}{\partial x_i} = \frac{d}{dt} \left( \frac{n}{c} \frac{1}{|\dot\x|} \dot x_i \right).
\end{equation}
It is easy to see that this is the same as Eq.~\eqref{eq1-for-v} that we found for the generic parameterization ($s=t$ and $v=c/n$ here).

\begin{acknowledgments}
The authors would like to acknowledge financial support from the research council of University of Tehran.  We also thank the anonymous referees for useful comments.
\end{acknowledgments}


\begin{thebibliography}{9}
\bibitem{geom-optics-approx} This is true as long as geometric optics is a good approximation, i.e., the wavelength of light is much smaller than the curvature scale of the surface.  We assume this to be the case throughout this paper.
\bibitem{nonholonomic} Constraints which depend on $\dot x$ are called non-holonomic.  When a non-holonomic constraint is not integrable to a holonomic one, the Euler-Lagrange equations fail to reproduce the correct equations of motion of classical mechanics.\cite{JS,F}  The failure of the variational formalism for classical mechanics may lead to the misconception that the problem is with the Euler-Lagrange equations.  However, the Euler-Lagrange equations provide the conditions for extremization of the action, whether or not the constraints are holonomic.\cite{GF,E}
\bibitem{domain-vs-endpoint} To avoid confusion, let us emphasize that the \textit{end-points} of the trajectories are still given by the same fixed points $\x_1$ an $\x_2$ of the previous parameterization.  It is only the \textit{domain} of the parameters like $s$, $t$, and $\ell$ that is different.
\end{thebibliography}
\end{document}